\documentclass[review]{article}
\usepackage{lineno,hyperref,latexsym,graphicx,amsmath}
\usepackage[font=small,labelfont=bf]{caption}

\usepackage[noend]{algorithmic}

\algsetup{indent=2em}
\newcounter{total_no_lines}

\newtheorem{thm}{Theorem}
\newtheorem{defin}{Definition}
\newtheorem{lemma}{Lemma}

\newtheorem{fact}{Fact}

\def\myqed{\hspace*{\fill}$\Box$\par\addvspace{\topsep}}

\title{Tree spanners of bounded degree graphs}

\author{
Ioannis Papoutsakis \\
Kastelli Pediados, Heraklion, Crete, Greece, 700 06
}

\begin{document}
\maketitle

\begin{abstract}
A  tree $t$-spanner of a graph $G$ is a spanning tree of $G$ such that the distance between pairs of vertices in
the tree is at most $t$ times their distance in $G$. Deciding tree $t$-spanner admissible graphs has been proved to be
tractable for $t<3$ and NP-complete for $t>3$, while the complexity status of this problem is unresolved when $t=3$.
For every $t>2$ and $b>0$, an efficient dynamic programming algorithm to decide tree $t$-spanner admissibility of graphs
with vertex degrees less than $b$ is presented. Only for $t=3$, the algorithm remains efficient, when graphs $G$
with degrees less than $b\log |V(G)|$ are examined.
\end{abstract}

{\small
{\bf Keywords}:
Tree spanner, Distance, Spanning tree, Efficient graph algorithm,  Bounded degree graph.}

\section{Introduction}

A $t$-spanner of a graph $G$ is a spanning subgraph of $G$, such that the distance between pairs of vertices in
the $t$-spanner is at most $t$ times their distance in $G$. Spanners, when they have a few edges, approximate
the distances in the graph, while they are sparse. Spanners of a graph that are trees attain the minimum number of edges a
spanner of the graph can have.
There are applications of spanners in a variety of areas, such as
distributed computing \cite{Awerbuch85,Peleg89},
communication networks \cite{PelUpf88,PelegReshef}, motion planning and
robotics \cite{Arikati96,Chew89}, phylogenetic analysis
\cite{Bandelt86} and in embedding finite
metric spaces in graphs approximately \cite{Rabinov98}.
In \cite{Pettie-Low-Dist-Span} it is mentioned that spanners have applications
in approximation algorithms for geometric spaces \cite{Narasimhanbook},
various approximation algorithms \cite{Fakcharoenphol}
and solving diagonally dominant linear systems \cite{Spielman}.

On one hand, in \cite{Bondy89,CaiCor95a,CaiThesis} an efficient algorithm
to decide tree 2-spanner admissible graphs is presented, where a method
to construct all the tree 2-spanners of a graph is also given. On the
other hand, in \cite{CaiCor95a,CaiThesis} it is proven that for
each $t\geq 4$ the problem to decide graphs that admit a tree $t$-spanner
is an NP-complete problem. The complexity status of the tree 3-spanner
problem is unresolved. In \cite{Fekete01}, for every $t$, an efficient algorithm to
determine whether a planar graph with bounded face length admits a tree $t$-spanner is
presented. In \cite{Fomin} the existence of an efficient (actually linear) algorithm for the tree spanner problem on
bounded degree graphs is shown, using a theorem of Logic; while it is mentioned that:
``It would be interesting to show that one could use tools that do not rely on
Courcelle’s theorem or Bodlaender’s algorithm to speed up practical implementations".
In this article, for every $t$, an efficient dynamic programming algorithm to
decide tree $t$-spanner admissibility of bounded degree graphs is presented
(theorem~\ref{tgeniko}).

Tree $t$-spanners ($t\geq 3$) have been studied for various families of
graphs. If a connected graph is a  cograph or a split graph or the
complement of a bipartite graph, then it admits a tree 3-spanner
\cite{CaiThesis}. Also, all convex bipartite graphs have a tree 3-spanner,
which can be constructed in linear time \cite{Venkatesan97}.
Efficient algorithms to recognize graphs that
admit a tree $3$-spanner have been developed for interval, permutation
and regular bipartite graphs \cite{Madanlal96}, planar graphs \cite{Fekete01},
directed path graphs \cite{Le99}, very strongly chordal graphs, 1-split graphs and
chordal graphs of diameter at most 2 \cite{Brandstadtchordal}. In \cite{Tree3-spannersofdiameteratmost5}
an efficient algorithm to decide if a graph admits a tree 3-spanner of diameter at most 5 is presented.
Moreover, every strongly chordal
graph admits a tree 4-spanner, which can be constructed in linear time
\cite{Brandst99}; note that, for each $t$, there is a connected chordal
graph that does not admit any tree $t$-spanner. The tree
$t$-spanner problem has been studied for small diameter chordal graphs \cite{Brandstadtchordal},
diametrically uniform graphs \cite{Manuel}, and
outerplanar graphs \cite{Narayanaswamy2015}. Approximation algorithms for the tree
$t$-spanner problem are presented in \cite{DraganK,PelegReshef}, where in \cite{DraganK} a new necessary condition
for a graph to have a tree t-spanner in terms of decomposition is also presented.

There are NP-completeness results for the tree $t$-spanner problem for families of graphs.
In \cite{Fekete01}, it is shown that it is NP-hard to determine the minimum $t$ for which a
planar graph admits a tree $t$-spanner. In \cite{Dragan2008}, it is proved that, for every $t \geq 4$, the problem of finding a
tree t-spanner is NP-complete on \mbox{$K_ 6$-minor-free} graphs.
 For any $t\geq4$, the tree $t$-spanner problem is NP-complete
on chordal graphs of diameter at most $t+1$, when $t$ is even, and of diameter at most $t+2$, when 
$t$ is odd \cite{Brandstadtchordal}; note that this refers to the diameter of the graph not to the diameter
of the spanner.  In \cite{Treespannersofsmalldiameter} it is shown that the
problem to determine whether a graph admits a tree $t$-spanner of
diameter at most $t+1$ is tractable, when $t\leq 3$, while it is an
NP-complete problem, when $t\geq 4$. This last result is used in  \cite{JCSS_inapprox} to hint
at the difficulty to approximate the minimum $t$ for which a graph
admits a tree $t$-spanner.

The tree 3-spanner problem is very interesting, since its complexity status is unresolved. In \cite{PhDthesis} it is
shown that only for $t=3$ the union of any two tree $t$-spanners of any given graph may contain big induced cycles but never
an odd induced cycle (other than a triangle); such unions are proved to be perfect graphs.
The algorithm presented in this article is efficient only for $t\leq 3$, when graphs with maximum degree $O(\log n)$
are considered, where $n(G)$ is the number of vertices of each graph $G$ (section~\ref{stequals3}).
The tree 3-spanner problem can be formulated as an integer
programming optimization problem. Constraints for such a formulation appear in \cite{PhDthesis}, providing certificates
of tree 3-spanner inadmissibility for some graphs.

\section{Definitions}
\label{sdefin}
In general, terminology of \cite{West} is used. If $G$ is a graph, then $V(G)$ is its {\em vertex set} and
$E(G)$ its {\em edge set}.
An {\em edge} between vertices $u,v\in G$ is denoted as $uv$. Also, $G\setminus\{uv\}$ is
the graph that remains when edge $uv$ is removed from $G$.
Let $v$ be a vertex of $G$, then $N_G(v)$ is the set of $G$ neighbors of $v$, while
$N_G[v]$ is $N_G(v)\cup \{v\}$; in this article, graphs do not have loop edges.
The {\em degree} of a vertex $v$ in $G$ is the number of edges of $G$ incident to $v$. Here,
$\Delta(G)$ is the maximum degree over the vertices of $G$.

Let $G$ and $H$ be two graphs. Then, $G\setminus H$ is graph $G$ without the vertices of $H$,
i.e. $V(G\setminus H)=V(G)\setminus V(H)$ and $E(G\setminus H)=\{uv\in E(G): u\not\in V(H)$ and $v\not\in V(H)\}$.
The {\em union} of $G$ and $H$, denoted as $G\cup H$, is the graph with vertex set $V(G)\cup V(H)$ and edge set
$E(G)\cup E(H)$. Similarly, the {\em intersection} of $G$ and $H$, denoted as $G\cap H$, is the graph with vertex set
$V(G)\cap V(H)$ and edge set $E(G)\cap E(H)$.
Additionally, $G[H]$ is the subgraph of $G$ {\em induced} by the vertices
of $H$, i.e. $G[H]$ contains all vertices in $V(G)\cap V(H)$ and all the edges of $G$ between vertices in $V(G)\cap V(H)$.
Note that the usual definition of induced subgraph refers to $H$ being a subgraph of $G$.

The $G$ distance between two vertices $u,v\in G$ is the length of a $u, v$ shortest path in $G$, while
it is infinity, when $u$ and $v$ are not connected in $G$.
The definition of a tree $t$-spanner follows.

\begin{defin}
A graph $T$ is a {\bf $t$-spanner} of a graph $G$ if and only if $T$ is a
subgraph of $G$ and, for every pair $u$ and $v$ of vertices of
$G$, if $u$ and $v$ are at distance $d$ from each other in $G$, then $u$
and $v$ are at distance at most $t\cdot d$ from each other in $T$. If $T$ is also a tree, then
$T$ is a {\bf tree $t$-spanner} of $G$.
\end{defin}

Note that in order to check that a spanning tree of a graph $G$
is a tree $t$-spanner of $G$, it suffices to examine pairs of vertices that are adjacent
in $G$ \cite{CaiThesis}. There is an additive version of a spanner as well
\cite{Kratsch98additivetree,Pettie-Low-Dist-Span}, which is not
studied in this article. In the algorithm and in the proofs, $r$-centers are frequently used.

\begin{defin}
Let $r$ be an integer. Vertex $v$ of a graph $G$ is an {\bf $r$-center} of $G$ 
if and only if for all vertices $u$ in $G$, the distance from $v$ to $u$ in $G$ is less than or equal to $r$.
\end{defin}

To refer to all the vertices near a central vertex, the notion of a
sphere is used.

\begin{defin}
Let $r$ be an integer and $v$ a vertex of a graph $G$. Then, the subgraph of $G$ induced by the vertices of
$G$ at distance less than or equal to $r$ from $v$ is the {\bf sphere} of $G$ with center $v$ and radius $r$;
it is denoted as \mbox{{\bf $(v,r)_G$-sphere}}.
\end{defin}

Obviously, $v$ is an $r$-center of a graph $G$, if and only if 
the $(v,r)_G$-sphere is equal to $G$.

Let $f, g$ be functions from the set of all graphs to the non negative integers. Then, $f$ is $O(g)$ if and only
if there are graph $G_0$ and integer $M$ such that $f(G)\leq M g(G)$ for every $G$ with $|V(G)|>|V(G_0)|$.
An algorithm that runs in polynomial time is called {\em efficient}.

\section{Description of the algorithm.}
In \cite{Master}, a characterization of tree $t$-spanner admissible graphs
in terms of decomposition states, generally speaking, that if a tree $t$-spanner
admissible graph $G$ does not have small diameter then it is the union of two tree
$t$-spanner admissible graphs whose intersection is a small diameter subgraph of $G$ (this result
requires further definitions to be stated exactly and it is not used in the proofs of this article).
So, it may be the case that, starting with small diameter subgraphs and adding on them
partial solutions of the remaining graph, a tree $t$-spanner of the whole graph is
built.

\begin{table}[tbp]
\begin{tt}
\begin{center}
\fbox{
\parbox{0.95\textwidth}{
{\bf Algorithm Find\_Tree\_spanner}($G$, $t$)\newline
{\bf Input:} A connected nonempty graph $G$ and an integer $t>1$.\newline
\begin{algorithmic}[1]
\STATE ${\cal A}_G^0=\emptyset$\label{l1-1}
\FOR{($k=1$ to $|V(G)|$)}\label{l1-2}
	\STATE ${\cal A}_G^k=\emptyset$\label{l1-3}
	\FOR{(vertex $v\in G$)}\label{l1-4}
		\STATE ${\cal S}_v = \{S\subseteq G: S$ is a tree $t$-spanner of $G[S]$ and
					$v$ is a $\lfloor\frac{t}{2}\rfloor$-center  of $S\}$\label{l1-5}
		\FOR{($S\in {\cal S}_v$)}\label{l1-6}
			\STATE $T^k_{v,S}=$ {\bf Find\_Subtree}($G$, $t$, $v$, $S$, $k$,${\cal A}_G^{k-1}$)\label{l1-7}
			\STATE ${\cal A}_G^k = {\cal A}_G^k \cup \{T^k_{v,S}\}$\label{l1-8}
			\STATE {\bf If} ($V(G)=V(T^k_{v,S})$) {\bf Return}($T^k_{v,S}$){\bf \}\}}\label{l1-9}
		\ENDFOR
	\ENDFOR
	\STATE {\bf Discard} ${\cal A}_G^{k-1}${\bf\}}\label{l1-10}
\ENDFOR
\STATE {\bf Return}($G$ does not admit a tree $t$-spanner.)\label{l1-11}
\setcounter{total_no_lines}{\value{ALC@line}}
\end{algorithmic}
}}
\end{center}
\end{tt}
\caption{Algorithm {\tt Find\_Tree\_spanner($G$, $t$)}. Procedure {\tt Find\_Subtree} is described in
table~\ref{tp}.}
\label{ta}
\end{table}

Algorithm {\tt Find\_Tree\_spanner} in table~\ref{ta} has as input a graph $G$
and an integer $t>1$.  Its output is a tree $t$-spanner of $G$ or a message that $G$ does not admit any
tree $t$-spanner. Being a dynamic programming algorithm, it grows partial solutions into final solutions starting from small 
subtrees of $G$. Obviously, each such subtree must be a tree $t$-spanner of the subgraph of $G$ induced by the vertices
of the subtree. All these subtrees are the first partial solutions of the dynamic programming method and are generated
by exhaustive search (first stage of the algorithm). Graphs of bounded degree have vertices of bounded neighborhoods;
therefore, this search for small subtrees is no harm. Note that the algorithm works for all input graphs but its efficiency
suffers when graphs of big degrees are examined.

\begin{table}[tbp]
\begin{tt}
\begin{center}
\framebox{
\noindent\parbox{0.95\textwidth}{
{\bf Procedure Find\_Subtree}($G$, $t$, $v$, $S$, $k$, ${\cal A}_G^{k-1}$)\newline
{\bf Input:} A graph $G$, an integer $t>1$, a vertex $v\in G$, a tree $t$-spanner $S$ of $G[S]$
with $\lfloor\frac{t}{2}\rfloor$-center $v$,
an integer $k\geq 1$, and a set ${\cal A}_G^{k-1}$ of subtrees of $G$.\newline
\begin{algorithmic}[1]
\setcounter{ALC@line}{\value{total_no_lines}}
\IF{($k=1$)}\label{l2-1}
	\STATE ${\cal Q}_{v,S}=\{Q\subseteq G: Q$ is a component of $G\setminus S\}$\label{l2-2}
				\COMMENT{static}
	\STATE {\bf Return}($S$){\bf\}}\label{l2-3}
\ELSE\label{l2-4}
	\STATE $T^k_{v,S}=T^{k-1}_{v,S}$  \COMMENT{$T^{k-1}_{v,S}$ is in ${\cal A}_G^{k-1}$}\label{l2-5}
	\FOR{(component $Q\in {\cal Q}_{v,S}$)}\label{l2-6}
		\FOR{($T^{k-1}_{u,R} \in {\cal A}_G^{k-1}$ such that $u\in N_S(v)$)}\label{l2-7}
			\STATE $T^k_{v,S,u,R,Q}=(T^{k-1}_{u,R}[Q\cup R]\cup S)\setminus((R\setminus S)\setminus Q)$\label{l2-8}
			\IF{($T^k_{v,S,u,R,Q}$ is a tree $t$-spanner of $G[Q\cup S]$)}\label{l2-9}
				\STATE $T^k_{v,S}=T^k_{v,S}\cup T^k_{v,S,u,R,Q}$\label{l2-10}
				\STATE ${\cal Q}_{v,S} = {\cal Q}_{v,S} \setminus \{Q\}$\label{l2-11}
				\STATE {\bf Break}{\bf\}\}\}}  \COMMENT{Stop search in ${\cal A}_G^{k-1}$}\label{l2-12}
			\ENDIF
		\ENDFOR
	\ENDFOR
	\STATE {\bf Return}($T^k_{v,S}$){\bf\}}\label{l2-13}
\ENDIF
\end{algorithmic}
}}
\end{center}
\end{tt}
\caption{Procedure {\tt Find\_Subtree($G$, $t$, $v$, $S$, $k$, ${\cal A}_G^{k-1}$)}. In line~\ref{l2-2}, variable ${\cal Q}_{v,S}$
has been declared as static; i.e. it is stored locally for later use, when the procedure is called again.}
\label{tp}
\end{table}

In each of the next stages of this dynamic programming algorithm, each partial solution is examined and, then, if possible, it is
incremented (procedure {\tt Find\_Subtree} in table~\ref{tp}).
The initial subtree of each partial solution (which was formed in the first stage) is its core.  Let $T^k_{v,S}$ be
a partial solution that is being examined. Removing the core of $T^k_{v,S}$, which is $S$, from $G$ creates
some components. Each such
component $Q$ that is not covered so far by $T^k_{v,S}$ is considered. The core of $T^k_{v,S}$ is put together with an
appropriate (based on $Q$) portion of nearby partial solutions; if the resulting graph is a tree $t$-spanner of the subgraph
of $G$ induced by the vertices of the resulting graph, then the partial solution under examination $T^k_{v,S}$ is incremented
by the resulting graph. This increment helps $T^k_{v,S}$ to cover $Q$. If $G$ admits a tree $t$-spanner, then some of
the partial solutions eventually cover $G$; if so, the algorithm outputs one of them (line~\ref{l1-9} of table~\ref{ta}).
Otherwise, $|V(G)|$ stages suffice to conclude that $G$ does not admit any tree $t$-spanner (line~\ref{l1-11} of
table~\ref{ta}). The description of the algorithm  in the two tables
has some details, which are explained in the following paragraphs.

Let us start with table~\ref{ta}.
Here, ${\cal A}_G^0$ is  set to $\emptyset$ (line~\ref{l1-1}) and its only use is to call a
procedure later on correctly.
To give motion to the process of growing partial solutions a main {\tt For} loop is used (line~\ref{l1-2}),
where variable $k$ is incremented
by 1 at the end of each stage, starting from 1. Set ${\cal A}_G^k$ is to store the progress on partial solutions and it is
initialized to $\emptyset$ (line~\ref{l1-3}); i.e. it is a set
whose elements are subtrees of $G$. The first stage (k=1) is different from the rest in not having previous
partial solutions to merge. First, it is necessary to pick names for the primary partial solutions. For each
vertex $v$ of $G$ a set ${\cal S}_v$ is formed (line~\ref{l1-5}). Each subgraph $S$ of $G$ that is a tree
$t$-spanner of $G[S]$ and has $v$ as a $\lfloor\frac{t}{2}\rfloor$-center becomes an element of ${\cal S}_v$.
This set ${\cal S}_v$ can be formed by exhaustively checking all the subtrees of the sphere of $G$ with
center $v$ and radius $\lfloor\frac{t}{2}\rfloor$ (see lemma~\ref{lsetsize}). Note that the computations to
form ${\cal S}_v$ can be done only for $k=1$. Then, for each member $S$ of ${\cal S}_v$ a partial solution is considered
under the name $T^1_{v,S}$. Second, each primary partial solution must be initialized (line~\ref{l1-7}). This is a job for procedure
{\tt Find\_Subtree}, which for $k=1$ returns $S$; i.e. $T^1_{v,S}=S$.
Of course, each newly formed primary partial
solution is stored in ${\cal A}_G^1$ (line~\ref{l1-8}). It may well be the case, when $G$ is a small graph, that some of these primary
solutions already spans $G$ (i.e. $V(G)=V(T^1_{v,S})$); then, a tree $t$-spanner of $G$ is found. This completes the
first stage of the main {\tt For} loop.

For $k>1$, partial solutions are merged if possible. Again, all partial solutions are considered one by one.
Procedure {\tt Find\_Subtree} in table~\ref{tp} receives as input (among others)
vertex $v$ and subtree $S\in{\cal S}_v$; these two determine the name of the partial solution under examination $T^k_{v,S}$,
where $k$ is just the number of the stage the algorithm is in.
It also receives as input all the partial solutions formed in the previous stage of the dynamic programming
method through set ${\cal A}_G^{k-1}$. Procedure {\tt Find\_Subtree} has saved locally the set of
components ${\cal Q}_{v,S}$ of $G\setminus S$, when it was called during the first stage of the main
algorithm ($k=1$). Set ${\cal Q}_{v,S}$ is a static variable; the content
of this set changes and these changes are remembered when the procedure is called again. Another way to
put it is that ${\cal Q}_{v,S}$ is a global variable, which is not lost each time the procedure ends.

The central set of operations of this dynamic programming algorithm is in procedure {\tt Find\_Subtree},
when $k>1$ (table~\ref{tp}, lines~\ref{l2-4} to~\ref{l2-13}) .
First, partial solution $T^k_{v,S}$ takes the value that it had in the previous stage, which had been
stored in ${\cal A}_G^{k-1}$; i.e $T^k_{v,S}=T^{k-1}_{v,S}$ (line~\ref{l2-5}).
Then, second, each component $Q$ in ${\cal Q}_{v,S}$ is examined to check if $T^k_{v,S}$ can be extended towards
$Q$ (line~\ref{l2-6}). Third, this extension will be done using other nearby partial solutions in ${\cal A}_G^{k-1}$.
For this, all partial solutions
in ${\cal A}_G^{k-1}$ that involve as central vertex a neighbor of $v$ in $S$ are considered, one at a time (line~\ref{l2-7});
the central vertex of partial solution $T^{k-1}_{u,R}$ is $u$.

\begin{figure}[htbp]
\begin{center}
\includegraphics[width=6cm]{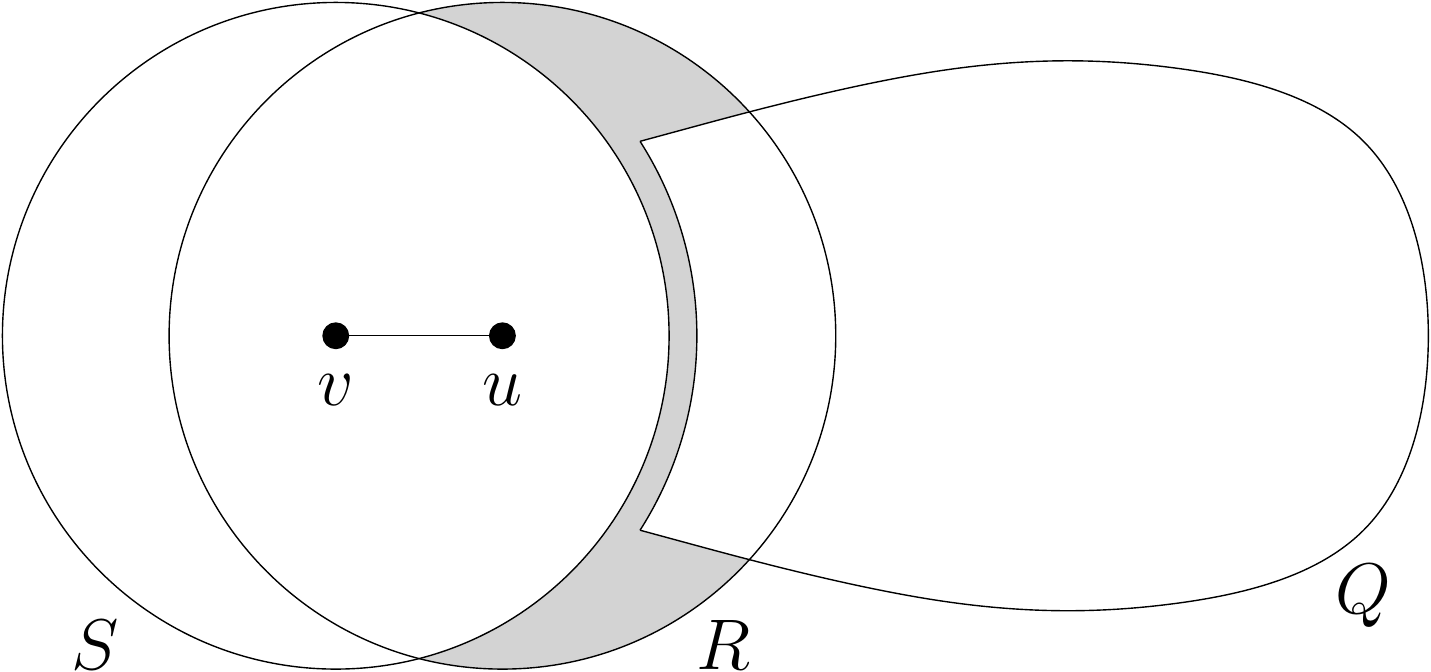}
\caption{The formation of auxiliary graph $T^k_{v,S,u,R,Q}$ in procedure {\tt Find\_Subtree} of 
table~\ref{tp}. The left circle is $S$, which is a tree
$t$-spanner of $G[S]$ and has $v$ as a $\lfloor\frac{t}{2}\rfloor$-center. Similarly, the right circle
is $R$, which is a tree $t$-spanner of $G[R]$ and has $u$ as a $\lfloor\frac{t}{2}\rfloor$-center;
here, $u$ is a neighbor of $v$ in $S$. Also, $Q$ is a component of $G\setminus S$. The
gray area is $(R\setminus S)\setminus Q$. The vertices that correspond to this gray area are removed
from $B=T^{k-1}_{u,R}[Q\cup R]\cup S$ to form the auxiliary graph.}
\label{fprocedurestep}
\end{center}
\end{figure}

Fourth, assume that the nearby partial solution  $T^{k-1}_{u,R}$ is considered when component $Q$ is examined. Then,
in line~\ref{l2-8}, an
auxiliary graph $T^k_{v,S,u,R,Q}$ is formed (see figure~\ref{fprocedurestep}). The fundamental part of this
auxiliary graph is $T^{k-1}_{u,R}[Q\cup R]$; i.e. the restriction of the considered nearby partial solution to the
component under examination (plus its core $R$). The intuition behind this operation is that $u$ may
be ``closer'' than $v$ to $Q$ and, therefore, $T^{k-1}_{u,R}$ may have covered
$Q$ in a previous stage of the dynamic programming method. For example, for $k=2$, input graph
$G$ may ``end'' towards the ``direction'' of edge $vu$; i.e. $V(Q)\subseteq V(R)$ and, therefore, $T^1_{u,R}$ covers $Q$.
In most of the cases $T^{k-1}_{u,R}$ does not contain all the vertices of $Q$; then, $T^{k-1}_{u,R}[Q\cup R]$ is still
meaningful, because of the slightly different than the usual definition of induced subgraph used in this article
(see section~\ref{sdefin}).
Having the foundation in hand, i.e. graph $T^{k-1}_{u,R}[Q\cup R]$, subtree $S$ is added to it; lets call for convenience
the resulting graph $B$ (i.e. $B=T^{k-1}_{u,R}[Q\cup R]\cup S$; here, $B$ is not used as a variable within
table~\ref{tp}).
Note that $B$ may not even be a tree, because $S\cup R$ may contain a cycle.
Since partial solution $T^k_{v,S}$ is to be incremented towards exactly $Q$, graph $(R\setminus S)\setminus Q$
(gray area in figure~\ref{fprocedurestep})
is removed\footnote{Due to the exhaustive search for primary partial solutions, there is always some $R'$ that
will do the job instead of $R$, such that $(R'\setminus S)\setminus Q=\emptyset$. But removing the gray area facilitates
the proof of correctness; this way more nearby partial solutions may help the partial solution under examination to grow.}
from $B$ to form the auxiliary graph $T^k_{v,S,u,R,Q}$ (see lemma~\ref{lgrayarea}).

Fifth, if auxiliary graph $T^k_{v,S,u,R,Q}$ is a tree $t$-spanner of $G[Q\cup S]$, then it is added to
the partial solution under examination; i.e. $T^k_{v,S}=T^k_{v,S}\cup T^k_{v,S,u,R,Q}$ (line~\ref{l2-10}). The result is again
a tree $t$-spanner of the subgraph of $G$ induced by the vertices of the result (see lemma~\ref{lmerikilysi}
and lemma~\ref{lunion}). Also, in this case, component $Q$ is removed from
$Q_{v,S}$ (line~\ref{l2-11}) and the search for nearby partial solutions to grow $T^k_{v,S}$ towards $Q$ stops
(line~\ref{l2-12}), since $T^k_{v,S}$ now covers $Q$. The execution continues after the {\tt If}
statement (line~\ref{l2-9}) with the
next component in $Q_{v,S}$. Note that auxiliary graph $T^k_{v,S,u,R,Q}$ can be discarded at this point.
These five steps complete the central set of operations of this dynamic
programming algorithm. Of course, after all components in $Q_{v,S}$ have been examined, procedure
{\tt Find\_Subtree} returns $T^k_{v,S}$ to the main program (line~\ref{l2-13}). 

A few comments on the algorithm follow. The algorithm works for $t=2$ as well, although there is an efficient algorithm
for this case \cite{Bondy89,CaiCor95a,CaiThesis}. Maintaining the various sets used in the algorithm is done using linked lists.
This way a {\tt For} loop on elements of a set retrieves sequentially all elements in a linked list.
Finally, procedure {\tt Find\_Subtree} doesn't need to check if its input is appropriate.

\section{Proof of correctness}

The following lemma is employed in various places within this section; when it is used in a proof, a footnote gives
the correspondence between the variable names in the proof and the names in its statement below. It
describes a basic property of spanners: vertices too far apart in a spanner of a graph cannot be adjacent in the graph.
The notion of a sphere has been defined in section~\ref{sdefin}

\begin{lemma}
Let $G$ be a graph, $T$ a tree $t$-spanner of $G$, and $x$ a vertex of $G$, where $t>1$. Let $X$ be the
$(x,\lfloor\frac{t}{2}\rfloor)_T$-sphere. Let $y$ be a $T$ neighbor of $x$
and let $T_y$ be the component of $T\setminus \{xy\}$ that contains $y$. Then, there is no edge of $G$ from a
vertex in $T_y\setminus X$ to a vertex in $(G\setminus T_y)\setminus X$.
\label{lcut_subtree}
\end{lemma}

{\em Proof}.
Assume, towards a contradiction, that there is an edge of $G$ between a vertex $p\in T_y\setminus X$ and
a vertex $q\in (G\setminus T_y)\setminus X$. Let $P_1$ be the $T$ path from $p$ to $x$. Then, all the
vertices of $P_1$ but $x$ are vertices of $T_y$. Also, the length of $P_1$ is strictly greater than $\lfloor\frac{t}{2}\rfloor$,
because $X$ contains all the vertices of $T$ at $T$ distance less than or equal to $\lfloor\frac{t}{2}\rfloor$
from $x$ and $p$ is out of $X$.
Let $P_2$ be the $T$ path from $q$ to $x$. Then, none of the
vertices of $P_2$ is a  vertex of $T_y$, because $q\not\in T_y$ and $x\not\in T_y$. Also, the length of $P_2$ is strictly
greater than $\lfloor\frac{t}{2}\rfloor$, because $X$ contains all the vertices of $T$ at $T$ distance less than or equal to 
$\lfloor\frac{t}{2}\rfloor$ from $x$ and
$q$ is out of $X$. Then, the $T$ path from $p$ to $q$ has length greater than or equal to $2\lfloor\frac{t}{2}\rfloor+2$; i.e. the
$T$ distance between the endpoints of edge $pq$ of $G$ is strictly greater than $t$. This is a contradiction to $T$ being
a tree $t$-spanner of $G$.\myqed

When growing a partial solution $T_1=T^k_{v,S}$ towards a component $Q$ of $G\setminus S$, by adding an
auxiliary graph $T_2=T^k_{v,S,u,R,Q}$ (line~\ref{l2-10} of table~\ref{tp}),
the result $T_1\cup T_2$  must be a tree $t$-spanner of $G[T_1\cup T_2]$;
the following lemma handles that. The vertices of forest $T_2\setminus T_1$  are the vertices of $Q$, while
the vertices of forest $T_1\setminus T_2$ are the vertices of the components of $G\setminus S$ that have already
been covered by $T_1$; since these two forests
correspond to such components, there is no edge of the input graph between them. The intersection
$T_1\cap T_2$ corresponds to the core (the initial value) of partial solution $T_1$, which is $S$.

\begin{lemma}
Let $G$ be a graph. Assume that $T_1$ is a tree $t$-spanner of $G[T_1]$ and that $T_2$ is a tree $t$-spanner of
$G[T_2]$. Also, assume that there is no edge of $G$ between a vertex in $T_1\setminus T_2$ and a vertex in
$T_2\setminus T_1$. If $T_1\cap T_2$ is a nonempty tree, then $T_1\cup T_2$ is a tree $t$-spanner of
$G[T_1\cup T_2]$.
\label{lunion}
\end{lemma}

{\em Proof}.
Let $T_{\cup}=T_1\cup T_2$ and $T_{\cap}=T_1\cap T_2$. First, it is proved that $T_{\cup}$ is a tree.
Let ${\cal Q}$ be the components of $T_{\cup}\setminus T_{\cap}$. Trivially, each  component in ${\cal Q}$ is either an
induced subgraph of $T_1\setminus T_2$ or an induced subgraph of $T_2\setminus T_1$.

Consider a component $Q$ in ${\cal Q}$. Obviously, $T_{\cup}$ is a connected graph ($T_{\cap}$ is nonempty).
So, there must be at least one
edge of $T_{\cup}$ between $Q$ and $T_{\cap}$. Without loss of generality, assume that $Q$ is an induced subgraph of
$T_1\setminus T_2$. So, all the edges of $T_{\cup}$ between $Q$ and $T_{\cap}$ belong to $T_1$.
Therefore, since $Q$ and $T_{\cap}$ are connected subgraphs of $T_1$ and $T_1$ is a tree, there is exactly one
edge of $T_{\cup}$ between $Q$ and $T_{\cap}$.
 
Here, $T_{\cap}$ is an induced subgraph of $T_{\cup}$, because any extra edge would form a cycle in $T_1$ or in $T_2$.
So, $T_{\cup}$ is a connected graph that consists of $|{\cal Q}|+1$ vertex disjoint trees plus  $|{\cal Q}|$ edges;
therefore, it is a tree.

Second, it is proved that $T_{\cup}$ is a $t$-spanner of $G[T_1\cup T_2]$. Consider an edge $vu$ of $G[T_1\cup T_2]$.
Then, $vu$  is an edge of $G[T_1]$ or an edge of $G[T_2]$, because
there is no edge of $G$ between a vertex in $T_1\setminus T_2$ and a vertex in $T_2\setminus T_1$. Assume,
without loss of generality, that $vu$ is an edge of  $G[T_1]$. Since $T_1$ is a tree $t$-spanner of $G[T_1]$, the distance
in $T_{\cup}$ between $v$ and $u$ is at most $t$.\myqed

Incrementing of partial solutions is done through a specific command within the algorithm. The following lemma
examines one by one the executions of this command and confirms that the incrementing is done properly. For
this, a double induction is used.

\begin{lemma}
Let $G$ be a graph and $t>1$ an integer. For every $v\in G$, for every $S\in{\cal S}_v$ and for every $k$ ($1\leq k \leq |V(G)|$),
$T^k_{v, S}$ is a tree $t$-spanner of $G[T^k_{v, S}]$, where ${\cal S}_v$ and $T^k_{v, S}$ are
constructed in algorithm {\tt Find\_Tree\_spanner} of table~\ref{ta} on input ($G$, $t$).
\label{lmerikilysi}
\end{lemma}

{\em Proof}.
Since $S\in{\cal S}_v$, $T^1_{v,S}$ belongs to ${\cal A}_G^1$ and it is equal to $S$,
which is a tree $t$-spanner of $G[S]$; so, the lemma holds for $k=1$.
Let $l_k$ be the total number of times $T^k_{v, S}$ is incremented through command
\begin{equation} \label{eincrement}
T^k_{v,S}=T^k_{v,S}\cup T^k_{v,S,u,R,Q}
\end{equation}
in line~\ref{l2-10} of table~\ref{tp}. Denote with $T^{k,l}_{v,S}$ the value of variable $T^k_{v,S}$, when
$T^k_{v, S}$ has been incremented $l$ times, through command~(\ref{eincrement}) above, where $2\leq k \leq |V(G)|$.
Then, $T^{2,0}_{v,S}=T^1_{v,S}$ and $T^{k,0}_{v,S}=T^{k-1,l_{k-1}}_{v,S}$, where $3\leq k \leq |V(G)|$
(line~\ref{l2-5} of table~\ref{tp}). Also,
$T^{k,l}_{v,S}=T^{k,l-1}_{v,S}\cup T^k_{v,S,u,R,Q}$, where $2\leq k \leq |V(G)|$ and $1\leq l\leq l_k$.

First, it is proved, by induction on $l$, that for each $k$ ($2\leq k \leq |V(G)|$) if $T^{k,0}_{v,S}$ is a tree $t$-spanner of $G[T^{k,0}_{v,S}]$, then $T^{k,l}_{v,S}$ is a tree $t$-spanner of $G[T^{k,l}_{v,S}]$, where $0\leq l\leq l_k$. The base
case ($l=0$) holds trivially. For the induction step ($1\leq l\leq l_k$), $T_1=T^{k,l-1}_{v, S}$ is incremented by
$T_2=T^k_{v,S,u,R,Q}$ for some $u$, $R$ and $Q$ to become $T^{k,l}_{v, S}$; i.e. $T^{k,l}_{v, S}=T_1\cup T_2$.
Then, $T_2$ must be a tree $t$-spanner
of $G[Q\cup S]$ (see condition of {\tt If} statement in line~\ref{l2-9} of table~\ref{tp}).
Also, $T_1$ is a tree $t$-spanner of $G[T_1]$, by induction hypothesis. The vertex set of $T_2$ is $V(S\cup Q)$.
Also, the vertex set of $T_1$ is the vertices of $S$ union the vertices of some components of
$G\setminus S$ other than $Q$,  because $Q$ has not been used to increment this partial solution
before. So, there is no edge of $G$ between a vertex in $T_1\setminus T_2$ and a vertex in
$T_2\setminus T_1$ and $V(T_1\cap T_2)=V(S)$. By construction of $T_2$, $S$ is a subtree of $T_2$. Also,
$T_1$ contains $S$, because $T_1$ contains $T^1_{v,S}$, which is equal to $S$. So, $T_1\cap T_2$ is $S$, which
is a nonempty tree. Here, $T^{k,l}_{v,S}=T_1\cup T_2$. Therefore, by lemma~\ref{lunion},
$T^{k,l}_{v,S}$ is a tree $t$-spanner of $G[T^{k,l}_{v,S}]$.

Second, the lemma is proved by induction on $k$; i.e. it is proved that $T^{k,l}_{v, S}$ is a tree $t$-spanner of
$G[T^{k,l}_{v, S}]$,
where $2\leq k \leq |V(G)|$ and $0\leq l\leq l_k$. For the base case, $T^{2,0}_{v,S}$ is equal to $T^1_{v,S}$,
which has been shown to be a tree $t$-spanner $G[T^1_{v,S}]$. So, by the first induction,
 $T^{2,l}_{v,S}$ is a tree $t$-spanner of $G[T^{2,l}_{v, S}]$, for $0\leq l\leq l_k$. For the induction step,
$T^{k,0}_{v,S}$ is equal to $T^{k-1,l_{k-1}}_{v,S}$, which, by induction hypothesis, is a tree $t$-spanner
$G[T^{k-1,l_{k-1}}_{v,S}]$. So, by the first induction,
$T^{k,l}_{v,S}$ is a tree $t$-spanner of $G[T^{k,l}_{v, S}]$, for $0\leq l\leq l_k$.\myqed

Assume that a partial solution $T^k_{v,S}$ is about to grow towards a component $W$ of $G\setminus S$ with the help
of a nearby partial solution $T'=T^{k-1}_{u,R}$. Then, the vertices of $R$ that are not in $S$ nor in $W$ are
not needed to grow $T^k_{v,S}$. The following lemma facilitates this process.
\begin{lemma}
Let $G$ be a graph and $T$ a tree $t$-spanner of $G$, where $t>1$. Let $v$ be a vertex of $G$ and let $S$ be the
$(v,\lfloor\frac{t}{2}\rfloor)_T$-sphere. Let $u$ be a $T$ neighbor of $v$ and let $R$ be the
$(u,\lfloor\frac{t}{2}\rfloor)_T$-sphere. Let $W$ be a component of $G\setminus S$. Let $T'$ be a
tree $t$-spanner of $G[T']$, such that $R\subseteq T'$. Let $L$ be $(R\setminus S)\setminus W$.
If $T'[W\cup R]\cup S$ is a tree $t$-spanner of $G[W\cup S\cup R]$, then $(T'[W\cup R]\cup S)\setminus L$
is a tree $t$-spanner of $G[W\cup S]$.
\label{lgrayarea}
\end{lemma}

{\em Proof}.
Obviously, $L$ corresponds to the gray area in figure~\ref{fprocedurestep}. Let $g$ be a vertex in $L$.
Then, $g\not\in S$ and $g\not\in W$. So, if there is an edge of $G$ from $g$ to a vertex in $W$, then $g$ must
be in $W$, a contradiction. Assume, towards a contradiction, that there is an edge $e$ of
$T'[W\cup R]\cup S$ from $g$ to a vertex in $S\cup R$,
such that $e\not\in E(S\cup R)$. Obviously, $S\cup R$ is a connected graph, because $\lfloor\frac{t}{2}\rfloor>0$.
Also, $R\subseteq T'$; so, all edges of $S\cup R$ are present in $T'[W\cup R]\cup S$. Therefore, there is
a path in $T'[W\cup R]\cup S$ between the endpoints of $e$ that avoids $e$ (note that $L\subseteq R$). This
is a contradiction to $T'[W\cup R]\cup S$ being a tree. So, since $L$ does not contain any vertex of $S$,
all edges of $T'[W\cup R]\cup S$ incident to $g$ must be edges of $R$ (it was proved earlier that there is no
edge of $G$ between $g$ and a vertex in $W$). Here, $g$ must be at distance
exactly $\lfloor\frac{t}{2}\rfloor$ from $u$, because $g\in R\setminus S$ and $S$ contains all the vertices at $T$ distance
less than or equal to $\lfloor\frac{t}{2}\rfloor-1$ from $u$. Therefore, $g$ is incident to only one edge of $T'[W\cup R]\cup S$.
So, removing $L$ from $T'[W\cup R]\cup S$ results to a tree $t$-spanner of $G[W\cup S].$\myqed

The main lemma in the proof of correctness of the algorithm follows. It guarantees that if the input graph
admits a tree $t$-spanner, then some partial solutions can grow during some stages of the algorithm. 
To break its proof into small parts, minor conclusions appear
as statements at the end of the paragraph that justifies them and are numbered equation like. Also,
intermediate conclusions appear as numbered facts. The lemma is proved
by induction on the number of stages of the algorithm (variable $k$ in the main {\tt For} loop at
table~\ref{ta}; see line~\ref{l1-2}). Note that the algorithm is ahead of the induction, in the sense
that for $k=2$ the algorithm starts merging primary partial solutions, while the induction considers such
merges for $k>\lfloor\frac{t}{2}\rfloor$, as one can see in the proof of fact~\ref{fspannerwhenempty} (for
$t\leq 3$, though, the algorithm and the induction are on the same page). The intuition behind the lemma is the
following. If a graph admits a tree $t$-spanner $T$, then there is a sphere $S$ of $T$ which is close to leaves of $T$, or
to picture it, say that $S$ is close to an end of $T$. Then, a nearby sphere $R$ does cover some of the leaves that
$S$ just misses. Here, $S$ corresponds to the partial solution that may grow, while $R$ corresponds to
a nearby partial solution $T^k_{u,R}$ that may help it grow. This picture is described formally by
fact~\ref{fspannerwhenempty}, where ${\cal H}=\emptyset$ means that $S$ is close to an end of $T$.
After the first steps of the induction, some partial solutions have grown. Assume now that $S$ is
not close to some end of $T$. Then, $S$ is 
nearby to a partial solution $T^k_{u,R}$, which is closer to that end of $T$ than $S$ is. The induction
hypothesis hints that, at some earlier stage, $T^k_{u,R}$ covered the part of the input graph
from $R$ to that end of $T$. So, $T^k_{u,R}$ may help (see fact~\ref{fspannerwhennotempty}) the partial solution
that corresponds to $S$ to grow.

\begin{lemma}
If $G$ admits a tree $t$-spanner $T$ ($t>1$) for which there exists vector ($k$, $v$, $S$, ${\cal W}$, $u$) such that:
\begin{enumerate}
\item $1\leq k\leq |V(G)|-1$, $v\in V(G)$,
\item $S$ is the $(v,\lfloor\frac{t}{2}\rfloor)_T$-sphere,
\item $u$ is a $T$ neighbor of $v$, $T_u$ is the component of $T\setminus\{uv\}$ that
contains $u$,
\item ${\cal W}=\{ X\subseteq G: X$ is a component of $G\setminus S$ and $V(X)\subseteq V(T_u)\}$, and
\item $v$ is a $k$-center of $T_u\cup S$,
\end{enumerate}
then algorithm {\tt Find\_Tree\_spanner} on input ($G$, $t$) (see table~\ref{ta}) returns a graph or for every
component $W\in {\cal W}$ there exists $R_W\subseteq G$ such that:
\begin{itemize}
\item $T^k_{u, R_W}$ is stored in ${\cal A}_G^k$ of algorithm {\tt Find\_Tree\_spanner} on input ($G$, $t$)
and
\item $T^{k+1}_{v,S,u,R_W,W}$ is a tree $t$-spanner of $G[W\cup S]$, where
$T^{k+1}_{v,S,u,R_W,W}$ is the graph $(T^k_{u,R_W}[W\cup R_W]\cup S)\setminus((R_W\setminus S)\setminus W)$
(see line~\ref{l2-8} of table~\ref{tp}, where such auxiliary graphs are constructed).
\end{itemize}
\label{lypodendro}
\end{lemma}

{\em Proof}.
Assume that algorithm {\tt Find\_Tree\_spanner} on input ($G$, $t$) does not return a graph; then all the stages of
the main {\tt For} loop of the algorithm are executed (line~\ref{l1-2} of table~\ref{ta}).
The lemma is proved by induction on $k$. For the {\bf base case}, $k\leq \lfloor\frac{t}{2}\rfloor$.
Here, $S$ is the subtree of $T$ that contains all the vertices of $T$ at $T$ distance less than or equal to $\lfloor\frac{t}{2}\rfloor$
from $v$. So, all the vertices in $\bigcup{\cal W}$ have to be at $T$ distance strictly greater than
$\lfloor\frac{t}{2}\rfloor$ from $v$ (each member of ${\cal W}$ is a component of $G\setminus S$).
Here, $v$ is a $k$-center of $T_u\cup S$; so,
all vertices in $\bigcup{\cal W}$ are at $T$ distance less than or equal to $k$ from $v$ (the vertex set of each component in
${\cal W}$ is subset of $T_u$). So, ${\cal W}=\emptyset$
and the lemma holds vacuously.

\begin{figure}[htbp]
\begin{center}
\includegraphics[width=7cm]{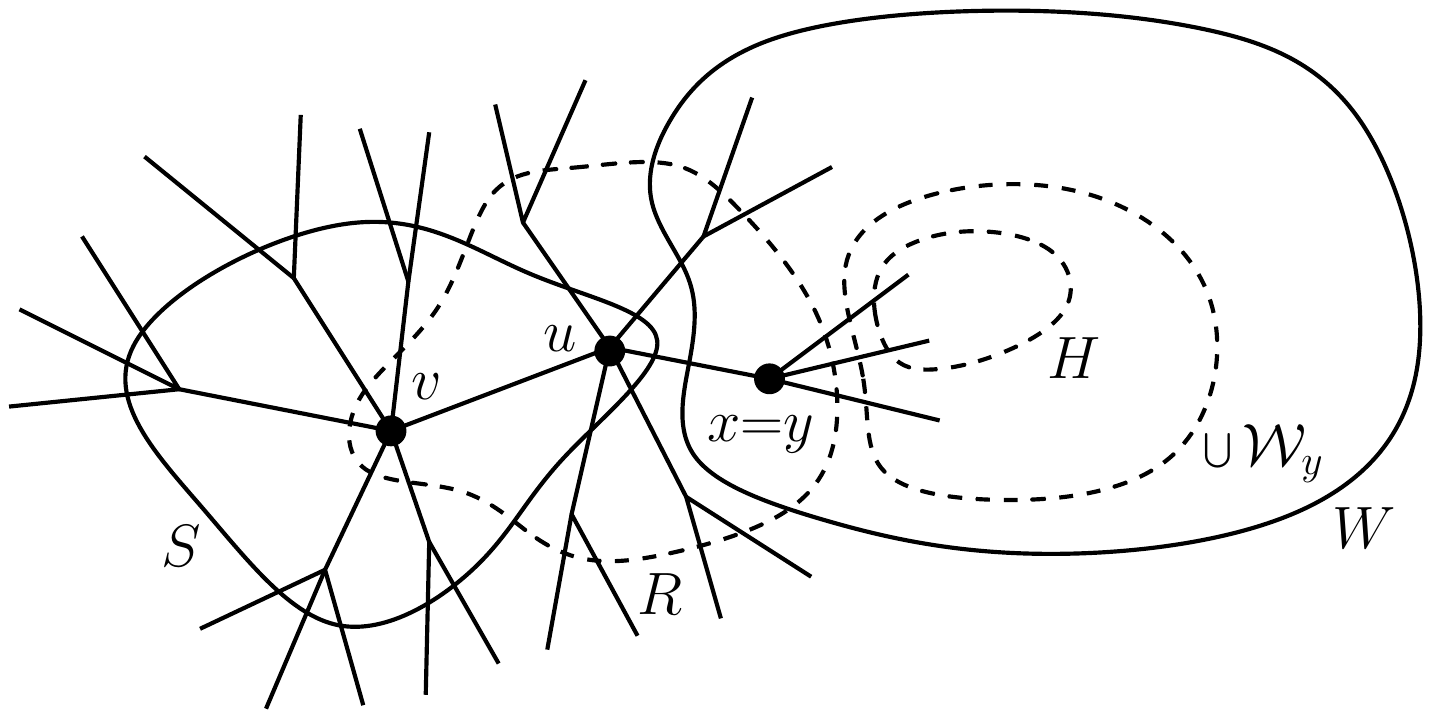}
\caption{The situation for $t=3$. Only edges of $T$ are shown. The dashed line sets concern vertex sets
involved in the induction hypothesis. Note that, when $t=3$, one can prove that for each $y\in X_W$, sets
${\cal W}_y$ and ${\cal W}'_y$ coincide.}
\label{fepanalipsi}
\end{center}
\end{figure}

For the {\bf induction step}, $\lfloor\frac{t}{2}\rfloor+1\leq k\leq |V(G)|-1$. Some definitions which
are used throughout the proof are introduced in this paragraph.
Let $W$ be a component in ${\cal W}$. Let $R$ be the $(u,\lfloor\frac{t}{2}\rfloor)_T$-sphere.
Let $X_W=V(R\cap W)$.
Let $Y_W=\{y\in N_T(u):$ there is a vertex $x\in X_W$ such that the $T$ path from $x$
to $u$ contains $y\}$. Here, $v$ is not in $Y_W$, because $V(W)\subseteq V(T_u)$. Note that, when
$t\leq3$, $X_W=Y_W$ (figure~\ref{fepanalipsi}). Now, for each $y\in Y_W$ let
$T_y$ be the component of $T\setminus\{uy\}$ that contains $y$. To make use of induction
hypothesis appropriate sets of components are defined. For each $y\in Y_W$ let
${\cal W}_y=\{X\subseteq G: X$ is a component of $G\setminus R$ and $V(X)\subseteq V(T_y)\}$. 
Since a tree $t$-spanner for $G[S\cup W]$ is to be constructed, only the components of
each ${\cal W}_y$ that fall within $W$ are of interest. So, for each $y\in Y_W$, let
${\cal W}'_y=\{H\in{\cal W}_y: H\subseteq W\}$. To refer to all these components,
define ${\cal H}=\bigcup_{y\in Y_W}{\cal W}'_y$.

\begin{figure}[htbp]
\begin{center}
\includegraphics[width=10cm]{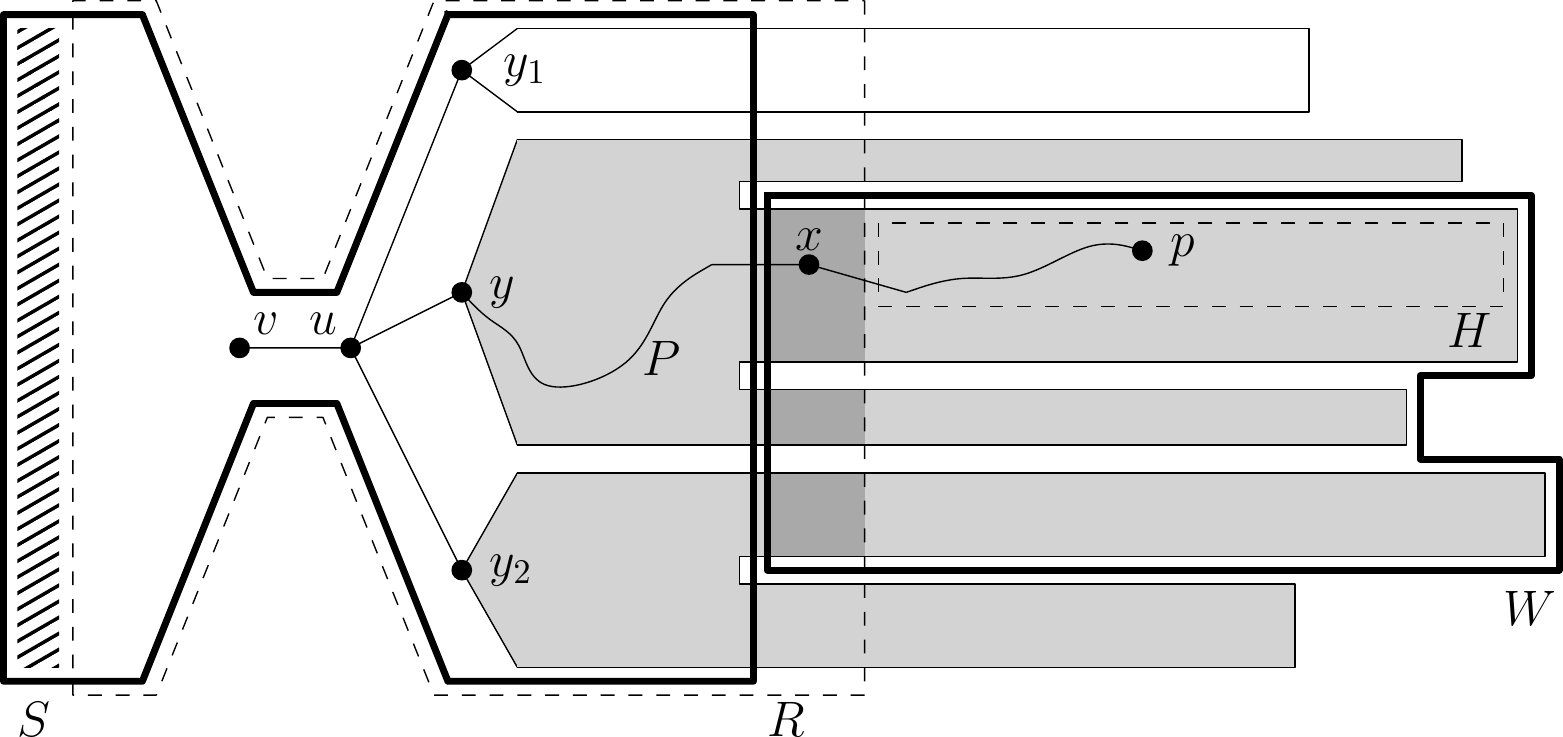}
\caption{The shapes with fat lines correspond to $S$ (left) and to $W$ (right). The shapes with dashed lines
correspond to $R$ (left) and to $H$ (right; also, $H$ may stand for $H_1$ as well, depending on  the context).
All the $T$ neighbors of $u$ are shown, namely, $v$, $y_1$, $y$, and $y_2$.
The dark gray area is set $X_W$. The subtrees of $T$ that correspond to $T_y$ and $T_{y_2}$ are shown gray (including
dark gray). Here, $y_1$ is not in $Y_W$. The gray area out of $R$ corresponds to $\bigcup_{y\in Y_W}(\bigcup{\cal W}_y)$,
while the gray area out of $R$ but within $W$ corresponds to $\bigcup {\cal H}$.
Here, $P$ is a $T$ path from $p$ to $v$; note that $P$ contains only one vertex in the dark gray area, namely $x$. Finally,
the hatched area is $S\setminus R$.}
\label{fsubtrees}
\end{center}
\end{figure}

\begin{figure}[htbp]
\begin{center}
\includegraphics[width=12cm]{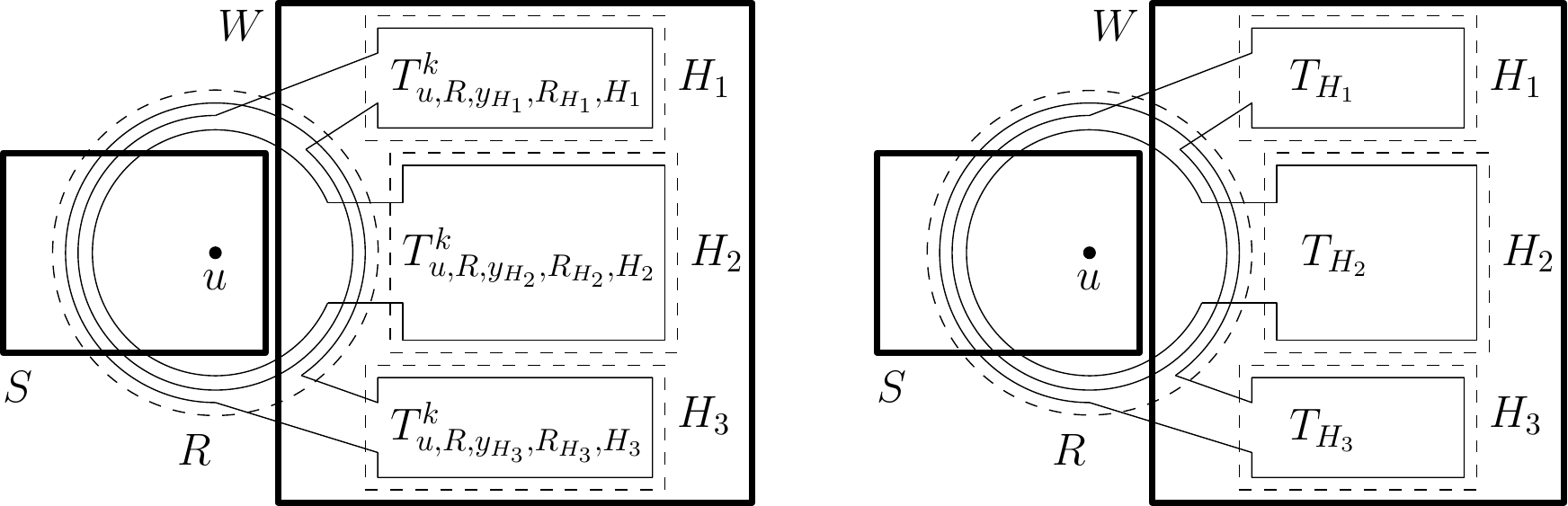}
\caption{Part of the situation in the proof of lemma~\ref{lypodendro}, as it is seen from two different angles.
The shapes with one circular end and one rectangular end correspond to auxiliary graphs.
The small rectangle is $S$ and the big rectagle is component $W$ of $G\setminus S$. The dashed circle is $R$; the dashed
rectangles are components of $G\setminus R$, which are the elements of ${\cal H}$ and are named as $H_1$, $H_2$, and $H_3$.
Here, $V(W)=V((W\cap R)\cup H_1\cup H_2\cup H_3)$ (see fact~\ref{fbiguinion}). The induction hypothesis dictates
that there are three auxiliary graphs, which are shown on the left hand side,
that can help partial solution $T^k_{u,R}$ to cover
$H_1$, $H_2$, and $H_3$. On the right hand side, three auxiliary graphs are also shown but are these that procedure
{\tt Find\_Subtree} (see table~\ref{tp}) actually picked to increment  partial solution $T^k_{u,R}$
towards $H_1$, $H_2$, and $H_3$. Note that the
pairwise intersection of the auxiliary graphs on each side is $R$; for example, $T_{H_1}\cap T_{H_2}=R$.
As the proof continues, $K=T_{H_1}\cup T_{H_2}\cup T_{H_3}$ becomes the
key element in proving the lemma, when ${\cal H}\not=\emptyset$.}
\label{falgochoice}
\end{center}
\end{figure}

A fundamental reason that the algorithm works is fact~\ref{fbiguinion}. It says that each component of
$G\setminus R$ falls either nicely into $W$ or completely out of $W$ and, therefore, the induction
hypothesis becomes useful (see figure~\ref{falgochoice}). On one hand, $X_W\subseteq V(W)$, by definition of $X_W$. Also,
For every $y\in Y_W$, every component in ${\cal W}'_y$ is a subgraph of $W$. Therefore,
\begin{equation} \label{eonedirection}
V(\bigcup{\cal H})\cup X_W\subseteq V(W)
\end{equation}
On the other hand, let $p$ be a vertex in $W$. Let $P$ be the $T$ path from $p$ to $v$ (see figure~\ref{fsubtrees}).
Since $V(W)\subseteq V(T_u)$, $P$ contains
$u$ and all its vertices but $v$ belong to $T_u$. All the vertices of $W$ are at $T$ distance strictly greater than
$\lfloor\frac{t}{2}\rfloor$ from $v$, by the definition of $S$ ($W$ is a component of $G\setminus S$). So, $P$ contains
exactly one vertex, say vertex $x$, at $T$ distance exactly $\lfloor\frac{t}{2}\rfloor+1$ from $v$ ($P$ is
a sub path of tree $T$). This means that $x$ is at $T$ distance exactly $\lfloor\frac{t}{2}\rfloor$ from $u$.
Therefore, $x\in R$ (note that $R$ contains all the vertices at $T$ distance less than or equal to
$\lfloor\frac{t}{2}\rfloor$ from $u$).
Also, all the vertices of $P$ from $x$ to $p$ are at $T$ distance strictly greater than $\lfloor\frac{t}{2}\rfloor$ from $v$; so,
there is a path in $G\setminus S$ from $p$ to $x$. But $p\in W$; so, $x\in W$ as well. Then, $x\in X_W$, because
$X_W=V(R\cap W)$. So,
\begin{equation} \label{episx}
p\in X_W, \text{when}\  p=x
\end{equation}
When $p\not=x$, $p$ is at $T$ distance strictly greater than $\lfloor\frac{t}{2}\rfloor$ from $u$, since
$x$ is at $T$ distance exactly $\lfloor\frac{t}{2}\rfloor$ from $u$;
so, $p\not\in R$. Therefore, $p$ is in a component, say component $H_1$, of
$G\setminus R$ (see figure~\ref{fsubtrees}). Assume, towards a contradiction, that $H_1\not\subseteq W$.
Here, $H_1$ is a component of $G\setminus R$, $W$ is a component of $G\setminus S$,
and $H_1\cap W\not=\emptyset$. So, since $H_1\not\subseteq W$, there must be an edge $e$ of $G$ from
a vertex in $H_1\cap W$ to a vertex in $S\setminus R$. As it can be seen in figure~\ref{fsubtrees}, the $T$ distance
between the endpoints of $e$ must be bigger than $t$, which contradicts to $T$ being a tree $t$-spanner of $G$; formally,
lemma~\ref{lcut_subtree} is employed.
Here, $S\setminus R$ is a subgraph of the component of $T\setminus\{uv\}$ that contains $v$ (easily seen by the
definitions of $S$ and $R$); call that component $T_v$.
So, $S\setminus R\subseteq T_v\setminus R$. Also, $V(W)\subseteq V(T_u)$ and
$H_1\subseteq G\setminus R$; so, $V(H_1\cap W)\subseteq V(T_u\setminus R$). But $T_v\cap T_u=\emptyset$; so,
$H_1\cap W\subseteq (G\setminus T_v)\setminus R$. Therefore, $e$ is an edge of $G$ from a vertex in $T_v\setminus R$
to a vertex in $(G\setminus T_v)\setminus R$; this is a contradiction to lemma\footnote{Vertex $u$ in the proof
corresponds to vertex $x$ in the lemma, $R$ to $X$, $v$ to $y$, and $T_v$ to $T_y$.}~\ref{lcut_subtree}.
So,
\begin{equation} \label{eHsubW}
H_1\subseteq W
\end{equation}
Let $y$ be the neighbor of $u$ in
$P$ towards $p$ (note that when $t\leq 3$, $x=y$); then, $y\in Y_W$ (since $x\in P$ and $x\in X_W$)
and $p\in T_y$. Assume, towards a contradiction, that there is a
vertex of $H_1$ out of $T_y$. Then, there must be an edge of $H_1$ (and of $G$ as well) from a vertex in $T_y\setminus R$
to a vertex in $(G\setminus T_y)\setminus R$, because $H_1$ is a component of $G\setminus R$. This is a contradiction to
lemma\footnote{Vertex $u$ in the proof
corresponds to vertex $x$ in the lemma and $R$ to $X$.}~\ref{lcut_subtree}. So,
\begin{equation} \label{eHsubTy}
V(H_1)\subseteq V(T_y)
\end{equation}
Here, $H_1$ is a component of $G\setminus R$, such that $V(H_1)\subseteq V(T_y)$ (statement~(\ref{eHsubTy}))  and
$H_1\subseteq W$ (statement~(\ref{eHsubW})). So, $H_1\in {\cal W}'_y$. Therefore, since
$y\in Y_W$, $H_1\in {\cal W}'_y$, and $p\in H_1$, it holds that:
\begin{equation} \label{episnotx}
p\in V(\bigcup{\cal H}), \text{when}\  p\not=x
\end{equation}
Since $p$ is just any vertex in $W$, from statements ~(\ref{eonedirection}),~(\ref{episx}), and~(\ref{episnotx}),
the following holds.
\begin{fact}
$V(\bigcup{\cal H})\cup X_W=V(W)$.
\label{fbiguinion}
\end{fact}

Sphere $R$ is going to be the required in the conclusion of the lemma $R_W$. Note that, by removing the
gray area as shown in figure~\ref{fprocedurestep}, sphere $R$ is suitable for every component
in ${\cal W}$, not just $W$. Trivially, $R$ is a tree $t$-spanner of $G[R]$, because it is a subtree
of $T$. Also, $u$ is a $\lfloor\frac{t}{2}\rfloor$-center of $R$, because $u$ is defined to be the center
of sphere $R$ of radius $\lfloor\frac{t}{2}\rfloor$. Therefore (line~\ref{l1-5} of table~\ref{ta}), $R\in {\cal S}_u$ and
the following holds (line~\ref{l1-8} of table~\ref{ta}).
\begin{fact}
Algorithm {\tt Find\_Tree\_spanner} on input ($G$, $t$) stores $T^k_{u, R}$ in ${\cal A}_G^k$
\label{fRinA}
\end{fact}

Next, it is proved that $T^{k+1}_{v,S,u,R,W}$ is a tree $t$-spanner of $G[W\cup S]$.
For this, {\bf two cases} are examined.

{\bf On one hand}, consider the case that
${\cal H}=\emptyset$. Then, an induction hypothesis cannot be used.
In this case, by fact~\ref{fbiguinion}, $V(W)=X_W$. But, $X_W\subseteq V(R)$, by definition.
So, $T^k_{u,R}[W\cup R]=T^k_{u,R}[R]$. But $R$ is equal to $T^1_{u, R}$, because, for $k=1$,
algorithm {\tt Find\_Tree\_spanner} on input ($G$, $t$) makes the call
{\tt Find\_Subtree}($G$, $t$, $u$, $R$, 1, $\emptyset$) and procedure {\tt Find\_Subtree} returns $R$.
Since partial solutions are never reduced during the algorithm, $R\subseteq T^k_{u,R}$. But
$T^k_{u, R}$ is a tree, because of lemma~\ref{lmerikilysi}. So, $T^k_{u,R}[W\cup R]=T^k_{u,R}[R]=R$, because
the subgraph of a tree induced by the vertices of a subtree is the subtree.
Therefore, $T^{k+1}_{v,S,u,R,W}$ is equal to $(R\cup S)\setminus ((R\setminus S)\setminus W)$.
By lemma~\ref{lgrayarea}, which ``clears" the gray area in figure~\ref{fprocedurestep},
it suffices to prove that $R\cup S$ is a tree $t$-spanner of $G[W\cup S\cup R]$.
Here, $R\cup S$ is a subtree of $T$, so it is a tree $t$-spanner of $G[S\cup R]$. But $V(W)\subseteq R$
in this case; so, $G[S\cup R]=G[W\cup S\cup R]$. Therefore, the following holds.
\begin{fact}
$T^{k+1}_{v,S,u,R,W}$ is a tree $t$-spanner of $G[W\cup S]$, when
${\cal H}=\emptyset$.
\label{fspannerwhenempty}
\end{fact}

{\bf On the other hand}, consider the case that ${\cal H}\not=\emptyset$. To prepare for
formulations of induction hypothesis, observe that $R$ is closer than $S$ to $W$. Formally,
all the vertices in $T_u$ are at $T$ distance less than or equal to $k$ from $v$, because
$v$ is a $k$-center of $T_u\cup S$. So, for each $y\in Y_W$, all vertices in $T_y$ are at $T$
distance less than or equal to $k-1$ from $u$. Also, all the vertices in $R$ are at $T$ distance less than
or equal to $\lfloor\frac{t}{2}\rfloor$ from
$u$. But in the induction step $k\geq \lfloor\frac{t}{2}\rfloor+1$; so, for each $y\in Y_W$, all the vertices in
$T_y\cup R$  are at $T$ distance less than or equal to $k-1$ from $u$ (note that $y\in R$, so $T_y\cup R$ is connected).
Therefore\footnote{Note that if $X$ is a connected subgraph of a tree $T$, then the $X$ distance between a pair of
vertices of $X$ is equal to the $T$ distance between this pair of vertices.},
\begin{equation}\label{ecenter}
\text{For each $y\in Y_W$, vertex $u$ is a $(k-1)$-center of $T_y\cup R$}
\end{equation}
Vector ($k-1$, $u$, $R$, ${\cal W}_y$, $y$)
satisfies the five conditions of the lemma for every $y\in Y_W$. To see this, first,
$1\leq\lfloor\frac{t}{2}\rfloor\leq k-1 \leq |V(G)|-2$ and
$u\in G$. Second, $R$ has been defined appropriately. Third, $y$ is a $T$ neighbor of
$u$ and $T_y$ has been defined appropriately. Fourth, ${\cal W}_y$ has been defined appropriately. Finally, fifth,
$u$ is a $(k-1)$-center of $T_y\cup R$ (statement~(\ref{ecenter})). Therefore,
since for every  $y\in Y_W$ the first coordinate of vector ($k-1$, $u$, $R$, ${\cal W}_y$, $y$) is strictly
less than $k$, the induction hypothesis states that the conclusion of the lemma holds.
Therefore, for every component in $\bigcup_{y\in Y_W}{\cal W}_y$ the two statements in the
conclusion of the lemma hold. But ${\cal H}\subseteq\bigcup_{y\in Y_W}{\cal W}_y$.
Let $H$ be a component in ${\cal H}$; then, $H\in {\cal W}'_{y_H}$ for some $y_H\in Y_W$.
Therefore, by the induction hypothesis, for $H$, there is $R_H\subseteq G$ such that (see figure~\ref{falgochoice}):
\begin{itemize}
\item $T^{k-1}_{y_H, R_H}$ is contained in ${\cal A}_G^{k-1}$ of algorithm {\tt Find\_Tree\_spanner($G$, $t$)} and
\item $T^k_{u, R, y_H,R_H,H}$ is a tree $t$-spanner of $G[H\cup R]$, where
$T^k_{u, R, y_H,R_H,H}$ is the graph
$(T^{k-1}_{y_H, R_H}[H\cup R_H]\cup R)\setminus((R_H\setminus R)\setminus H)$.
\end{itemize}

Therefore, when procedure {\tt Find\_Subtree} is called (line~\ref{l1-7} of table~\ref{ta}) with input
($G$, $t$, $u$, $R$, $k$, ${\cal A}_G^{k-1}$) there is at least one
entry in ${\cal A}_G^{k-1}$ to satisfy the condition (line~\ref{l2-9} of table~\ref{tp})  for incrementing $T^k_{u, R}$
towards $H$; note that $y_H\in Y_W\subseteq N_T(u)=N_R(u)$.
So (see figure~\ref{falgochoice}), for some $l_H\leq k$, $T^{l_H}_{u, R}$ is incremented towards $H$ and assume that
the procedure picks $T_H=T^{l_H}_{u, R, p_H, R(p_H,H),H}$ to do so, where $p_H$ is some $R$ neighbor
of $u$ and $T^{l_H-1}_{p_H,R(p_H,H)}$ belongs to ${\cal A}^{l_H-1}_G$; i.e. $T_H$ is a tree
$t$-spanner of $G[H\cup R]$ and the command $T^{l_H}_{u, R}=T^{l_H}_{u, R}\cup T_H$ is executed
(lines~\ref{l2-9} and~\ref{l2-10} of table~\ref{tp}).
Since $T_H\subseteq T^{l_H}_{u, R}$ and $ T^{l_H}_{u, R}\subseteq  T^k_{u, R}$,
auxiliary graph $T_H$ is a subgraph of $T^k_{u, R}$. All these
auxiliary graphs that correspond to each $H\in{\cal H}$ (which have been used by procedure
{\tt Find\_Subtree} to construct a part of $T^k_{u, R}$) are put
together in one graph $K$. The following holds.
\begin{equation}
\text{Let}\ K=\bigcup_{H\in {\cal H}}T_H.\ \text{Then,}\ K\subseteq T^k_{u, R}
\label{einductionhypothesis}
\end{equation}

For each $H\in {\cal H}$ auxiliary graph $T_H$ is a connected graph,
because it is a tree $t$-spanner of $G[H\cup R]$. Also, all these auxiliary graphs
share the vertices of $R$.
So, $K$ is a connected subgraph of $T^k_{u,R}$ (statement~(\ref{einductionhypothesis})).
But $T^k_{u, R}$ is a tree $t$-spanner of $G[T^k_{u, R}]$,
because of lemma~\ref{lmerikilysi}. So, $K$ is a tree $t$-spanner of $G[K]$.
Also, $K$ coincides with $T^k_{u,R}[K]$, because it is a subtree of tree $T^k_{u,R}$.
Here, $V(K)=V((\bigcup{\cal H})\cup R)$,
again because for each $H\in {\cal H}$ auxiliary graph $T_H$ is a tree 
$t$-spanner of $G[H\cup R]$. Therefore, by fact~\ref{fbiguinion}, $V(K)=V(W\cup R)$, because
$X_W\subseteq V(R)$. So, $K$ is a tree $t$-spanner of $G[W\cup R]$ and coincides with
$T^k_{u,R}[W\cup R]$. The following holds.
\begin{equation}
\text{$K=T^k_{u,R}[W\cup R]$ is a tree $t$-spanner of $G[W\cup R]$}
\label{eWcupR}
\end{equation}

Set $T_2=R\cup S$. Here, $T_2$ is a connected subgraph of $T$; so, $T_2$ is a tree $t$-spanner of $G[T_2]$.
Lemma~\ref{lunion} will be used to prove that $K\cup T_2$ is a tree $t$-spanner of $G[K\cup T_2]$, so the
additional requirements for $K$ and $T_2$ are shown\footnote{Here, $K$ corresponds to $T_1$ of the lemma.}.
Assume, towards a contradiction, that there is an
edge $e'$ of $G$ between a vertex in $K\setminus T_2$ and a vertex in $T_2\setminus K$ (a similar approach was
taken in the proof of statement~(\ref{eHsubW}), where tree $T_v$ was defined to be the component of
$T\setminus\{uv\}$ that contains $v$). Here, $V(K\setminus T_2)$ is a subset of $V(T_u\setminus R)$,
because $V(W)\subseteq V(T_u)$ (here $V(K)=V(W\cup R)$ and $R\subseteq T_2$). So, since
$T_v\cap T_u=\emptyset$, it holds that $V(K\setminus T_2)\subseteq V((G\setminus T_v)\setminus R)$.
Also, $T_2\setminus K$ is equal to $S\setminus R$, because $W$ being a component of
$G\setminus S$ avoids $S$. Clearly, $V(S\setminus R)$ is a subset of $V(T_v\setminus R)$. Therefore, the
existence of edge $e'$ is a contradiction to lemma\footnote{Vertex $u$ in the proof corresponds to vertex $x$
in the lemma, $R$ to $X$, $v$ to $y$, and $T_v$ to $T_y$.}~\ref{lcut_subtree}. Here, $K\cap T_2$ is
equal to $R$, which is a nonempty tree. Therefore, by lemma~\ref{lunion}, $K\cup T_2$ is a tree $t$-spanner
of $G[K\cup T_2]$. But $K\cup T_2$ is equal to $K\cup S$, because $R\subseteq K$. Therefore (see statement~(\ref{eWcupR})),
the following holds.
\begin{equation}
\text{$K\cup S=T^k_{u,R}[W\cup R]\cup S$ is a tree $t$-spanner of $G[W\cup R\cup S]$}
\label{eWcupRcupS}
\end{equation}

Here, $T^{k+1}_{v,S,u,R,W}$ is equal to $(T^k_{u,R}[W\cup R]\cup S)\setminus ((R\setminus S)\setminus W)$;
so, it remains to remove the gray area in figure~\ref{fprocedurestep}.
Then, by statement~(\ref{eWcupRcupS}) and lemma~\ref{lgrayarea}, the following holds.
\begin{fact}
$T^{k+1}_{v,S,u,R,W}$  is a tree $t$-spanner of $G[W\cup S]$, when
${\cal H}\not=\emptyset$.
\label{fspannerwhennotempty}
\end{fact}

By facts~\ref{fRinA},~\ref{fspannerwhenempty}, and~\ref{fspannerwhennotempty} the lemma holds.\myqed

To explain the time complexity of the algorithm, when graphs of bounded degree are inputs, the following lemma
binds the size of various sets used in {\tt For} loops by  functions of the maximum degree of the input graph.

\begin{lemma}
Let $G$ be a connected graph of maximum degree $\Delta$ and $t>1$ an integer. Then,
\begin{enumerate}
\item $|{\cal S}_v|\leq 2^{\Delta^{2+\lfloor\frac{t}{2}\rfloor}}$, for every vertex $v$ of $G$,
\item $|{\cal Q}_{v,S}|\leq \Delta^{2+\lfloor\frac{t}{2}\rfloor}+\Delta$, for every vertex $v$ of $G$
and for every $S$ in ${\cal S}_v$,
\item $|{\cal A}_G^{k-1}|\leq |V(G)| \max_{x\in V(G)}|{\cal S}_x|$, for every $k$ ($2\leq k \leq |V(G)|$),
\item the number of $T^{k-1}_{u,R} \in {\cal A}_G^{k-1}$ such that $u\in N_S(v)$ is at most
$\Delta \max_{x\in V(G)} |{\cal S}_x|$, for every vertex $v$ of $G$,
for every $S$ in ${\cal S}_v$, and for every $k$ ($2\leq k \leq |V(G)|$),
\end{enumerate}
where, ${\cal S}_v$, ${\cal Q}_{v,S}$, ${\cal A}_G^{k-1}$, and $T^{k-1}_{u,R}$ are constructed in algorithm
{\tt Find\_Tree\_spanner} of tables~\ref{ta} and~\ref{tp}
on input ($G$, $t$).
\label{lsetsize}
\end{lemma}

{\em Proof}.
First, let $G_v$ be the sphere of $G$ with center $v$ and radius $\lfloor\frac{t}{2}\rfloor$. Let $S$ be a member
of ${\cal S}_v$. Since $v$ is a $\lfloor\frac{t}{2}\rfloor$-center of $S$ (line~\ref{l1-5} of table~\ref{ta}),
$S$ must be a subgraph of $G_v$.
Let $L^i$ denote the number of vertices at $G$ distance exactly $i$ from $v$. Then $L^0=1$. Also, $L^i\leq \Delta L^{i-1}$,
for $i\geq 1$, because $G$ has maximum degree $\Delta$ and each vertex at $G$ distance
$i$ from $v$ must be adjacent to a vertex at $G$ distance $i-1$ from $v$. So, $G_v$ has at most
$\sum_{i=0}^{i=\lfloor\frac{t}{2}\rfloor}L^i$ vertices. So, simply, $G_v$ has  at most
$\Delta^{1+\lfloor\frac{t}{2}\rfloor}+1$ vertices\footnote{This includes the $\Delta\leq 1$ cases,
because, then, $G_v$ is the one vertex graph ($\Delta=0$) or has at most two vertices ($\Delta=1$).}.
Therefore, since $G$ has maximum degree $\Delta$, $G_v$ has at most $\Delta |V(G_v)|/2$
edges. So, simply, $G_v$ has at most
$\Delta^{2+\lfloor\frac{t}{2}\rfloor}$ edges. At most $|V(G_v)|-1$ edges can participate in $S$. So, roughly, considering the
power set of $E(G_v)$, the number of elements in ${\cal S}_v$ can be at most $2^{\Delta^{2+\lfloor\frac{t}{2}\rfloor}}$.

Second, any subtree $S$ in ${\cal S}_v$ can have at most $|V(G_v)|$ vertices. Then, as shown earlier, $S$ can 
have at most $\Delta^{1+\lfloor\frac{t}{2}\rfloor}+1$ vertices.
The number of edges of $G$ with one endpoint in $S$ and the other
out of $S$ can be at most $\Delta|V(S)|$, since $G$ has maximum degree $\Delta$. Then, since $G$ is
connected,\footnote{The algorithm works for disconnected graphs as well without increasing its time complexity.
Here is the only place the connectedness of the input graph is used. This facilitates the calculations for the running
time of the algorithm.}
the number of components of $G\setminus S$ is at most $\Delta^{2+\lfloor\frac{t}{2}\rfloor}+\Delta$.

Third, for each vertex $x$ of $G$, $|{\cal S}_x|$ partial solutions are in ${\cal A}_G^{k-1}$, where $2\leq k \leq |V(G)|$.
So, $|{\cal A}_G^{k-1}|\leq |V(G)| \max_{x\in V(G)}|{\cal S}_x|$.

Fourth, let $v$ be a vertex of $G$ and $S$ a member of ${\cal S}_v$. Then, since $G$ has maximum degree $\Delta$,
$v$ has at most $\Delta$ neighbors in $S$. But each vertex $u$ of $G$ is a central vertex of $|{\cal S}_u|$ partial
solutions. Therefore, the number of $T^{k-1}_{u,R} \in {\cal A}_G^{k-1}$ such that $u\in N_S(v)$ is at most
$\Delta \max_{x\in V(G)}|{\cal S}_x|$, where $2\leq k \leq |V(G)|$.
\myqed

\begin{thm}
Let $b$, $t$ be positive integers. There is an efficient algorithm to decide whether any graph $G$ with $\Delta(G)\leq b$
admits a tree $t$-spanner.
\label{tgeniko}
\end{thm}

{\em Proof}.
If $t=1$, then a graph admits a tree 1-spanner if and only if it is a tree.
The empty graph admits a tree $t$-spanner and a disconnected graph cannot admit a tree $t$-spanner.
So, it remains to check nonempty connected graphs for $t>1$. For this, the algorithm described in this article
is employed and it is proved that a nonempty connected graph $G$ admits a tree $t$-spanner if and only if
algorithm {\tt Find\_Tree\_spanner} on input ($G$, $t$) returns a graph, where $t>1$.

For the sufficiency proof, assume that algorithm {\tt Find\_Tree\_spanner} on input
($G$, $t$) returns $T^k_{v,S}$ (line~\ref{l1-9} of table~\ref{ta}).
Then, $V(G)=V(T^k_{v,S})$. But $T^k_{v,S}$ is a tree $t$-spanner
of $G[T^k_{v,S}]$, because of lemma~\ref{lmerikilysi}. Therefore, $G$ admits a tree $t$-spanner.

For the necessity proof, assume that $G$ admits a tree $t$-spanner $T$. Let $v$ be a vertex of $G$ and
$S$ be the $(v,\lfloor\frac{t}{2}\rfloor)_T$-sphere.
Then, algorithm {\tt Find\_Tree\_spanner} on input ($G$, $t$) calls
procedure {\tt Find\_Subtree} with parameters ($G$, $t$, $v$, $S$, $1$, ${\cal A}_G^0$)
(line~\ref{l1-7} of table~\ref{ta}); note that
this happens even when $G$ is the one vertex graph. Then, procedure {\tt Find\_Subtree} returns $S$
(line~\ref{l2-3} of table~\ref{tp}) which becomes $T^1_{v,S}$.

On one hand, consider the case that ${\cal Q}_{v,S}$ is empty (line~\ref{l2-2} of table~\ref{tp}).
Then, $G$ contains no vertices out of $S$; so, the algorithm returns $T^1_{v,S}$.

On the other hand, consider the case that ${\cal Q}_{v,S}$  is not empty. Assume, towards a contradiction,
that algorithm {\tt Find\_Tree\_spanner} on input ($G$, $t$) does not return a graph. Then,
for $k=|V(G)|$ the algorithm {\tt Find\_Tree\_spanner} on input ($G$, $t$) does not return
$T^k_{v,S}$ (line~\ref{l1-9} of table~\ref{ta}).
This means that $T^k_{v,S}$ does not contain the vertex set of some component
$W$ in ${\cal Q}_{v,S}$. This happens because (lines~\ref{l2-6} to~\ref{l2-10} of table~\ref{tp})
there is no vertex $x\in N_S(v)$ and $R_W\subseteq G$,
such that $T^{k-1}_{x, R_W}$ is contained in set ${\cal A}_G^{k-1}$ and
$T^k_{v,S,x,R_W,W}$ is a tree $t$-spanner of $G[W\cup S]$, where
$T^k_{v,S,x,R_W,W}=(T^{k-1}_{x,R_W}[W\cup R_W]\cup S)\setminus((R_W\setminus S)\setminus W)$.
But this is a contradiction to lemma~\ref{lypodendro}; to see this, its five conditions are examined.
First, $1\leq k-1\leq |V(G)|-1$, because
$|V(G)|>1$ in this case; also, $v\in V(G)$. Second, $S$ is the $(v,\lfloor\frac{t}{2}\rfloor)_T$-sphere.
Third, let $u$ be a $T$ neighbor of $v$ that is in a $T$ path from $W$ to $v$;
also, let $T_u$ be the component of $T\setminus\{uv\}$ that contains $u$. Fourth,
let ${\cal W}$ be the set $\{ X\subseteq G: X$ is a component of $G\setminus S$ and $V(X)\subseteq V(T_u)\}$.
Finally, fifth, $v$ is a $(k-1)$-center of $T_u\cup S$, even when $G$ is a path and $v$ an end vertex ($k=|V(G)|$
and $T_u\cup S$ is connected, because $u\in S$).
Therefore, vector ($k-1$, $v$, $S$, ${\cal W}$, $u$) satisfies the conditions of lemma~\ref{lypodendro}.

It suffices to prove that $W\in {\cal W}$. By definition of vertex $u$, at least one vertex of $W$
belongs to $T_u$. Assume, towards a contradiction, that there is a vertex of $W$ which
is not in $T_u$. Then, since $W$ is a component of $G\setminus S$, there must be an edge of $G$ from a vertex in
$T_u\setminus S$ to a vertex in $(G\setminus T_u)\setminus S$. This is a contradiction to lemma\footnote{Vertex
$v$ in the proof corresponds to vertex $x$ in the lemma, $S$ to $X$, $u$ to $y$, and $T_u$ to $T_y$.}~\ref{lcut_subtree}.
Hence, $W\in {\cal W}$. Therefore, by lemma~\ref{lypodendro} there is such a vertex $x$,
namely $u$, and such an $R_W$; a contradiction.

It remains to prove that the algorithm runs in polynomial time, when bounded degree graphs are considered.
Let $n(G)$ be the number of vertices in $G$. Then, checking if the input to algorithm {\tt Find\_Tree\_spanner} is
a connected nonempty graph $G$ with $\Delta(G)\leq b$ and $t>1$ takes $O(n)$ time.
For every vertex $v\in G$, $|{\cal S}_v|$ is $O(1)$,
because $\Delta(G)$ is bounded by constant $b$ and $t$ is a constant (see lemma~\ref{lsetsize}). So,
procedure {\tt Find\_Subtree} in table~\ref{tp} is called $O(n^2)$ times.

Consider procedure {\tt Find\_Subtree} on input ($G$, $t$, $v$, $S$, $k$, ${\cal A}_G^{k-1}$).
The construction of ${\cal Q}_{v,S}$ takes $O(n)$ time ($G$ has a linear number of edges) and it is done only when $k=1$. 
The commands in lines~\ref{l2-5} to~\ref{l2-11} of procedure {\tt Find\_Subtree} in table~\ref{tp} are executed
when $k>1$ and are examined one by one in this paragraph. The number of partial solutions formed in the previous stage
$|{\cal A}_G^{k-1}|$ is $O(n)$ (lemma~\ref{lsetsize}), because $|{\cal S}_x|$ is $O(1)$ for each vertex $x\in G$.
Finding $T^{k-1}_{v,S}$ in ${\cal A}_G^{k-1}$ and doing the assignment $T^k_{v,S}=T^{k-1}_{v,S}$ (line~\ref{l2-5})
takes $O(n)$ time, because $|{\cal A}_G^{k-1}|$ is $O(n)$. Also, by lemma~\ref{lsetsize}, $|{\cal Q}_{v,S}|$ is $O(1)$,
because $\Delta(G)$ is bounded by constant $b$ and $t$ is a constant (line~\ref{l2-6}). Next,  it takes $O(n)$ time to
check sequentially all elements of ${\cal A}_G^{k-1}$, because $|{\cal A}_G^{k-1}|$ is $O(n)$. Though, (line~\ref{l2-7})
the number of $T^{k-1}_{u,R} \in {\cal A}_G^{k-1}$ such that $u\in N_S(v)$ is $O(1)$ (lemma~\ref{lsetsize}),
because $\Delta(G)$ is bounded by constant $b$ and $|{\cal S}_x|$ is $O(1)$ for each vertex $x\in G$.
Construction of $T^{k-1}_{u,R}[Q\cup R]$ in line~\ref{l2-8} takes $O(n)$ time. To check whether $T^k_{v,S,u,R,Q}$ is a tree
$t$-spanner of $G[Q\cup S]$ in line~\ref{l2-9} takes $O(n)$ time, because $\Delta(G)$ is bounded by constant $b$ and,
therefore, $G[Q\cup S]$ has a linear number of edges. Constructing the union of $T^k_{v,S}$ and
$T^k_{v,S,u,R,Q}$ in line~\ref{l2-10} takes $O(n)$ time. Finally, removing $Q$ from ${\cal Q}_{v,S}$ in line~\ref{l2-11}
can be done in linear time. Therefore, each call of procedure {\tt Find\_Subtree} takes $O(n)$ time.

Returning back to algorithm {\tt Find\_Tree\_spanner} (lines~\ref{l1-7} to~\ref{l1-9} of table~\ref{ta}),
inserting the output of procedure {\tt Find\_Subtree} in
${\cal A}_G^{k}$ takes $O(n)$ time. Next, checking whether $V(G)=V(T^k_{v,S})$ and output $T^k_{v,S}$
take $O(n)$ time. All these three commands are executed $O(n^2)$ times, i.e. as many times as
procedure {\tt Find\_Subtree} is called. Therefore, algorithm {\tt Find\_Tree\_spanner} takes $O(n^3)$
time and it is efficient.\myqed

\section{The $t=3$ case}
\label{stequals3}
Let $G$ be a graph. When $t=3$, for every $v$ in $G$, each $S$ in ${\cal S}_v$ must have $v$ as a
$1$-center, because $\lfloor\frac{t}{2}\rfloor=1$ (line~\ref{l1-5}
of table~\ref{ta}). This means that $S$ is a tree with central vertex
$v$ and all its remaining vertices are leaves. If the maximum degree of $G$ is $\Delta$, then $S$ can
have up to $\Delta$ leaves. So, ${\cal S}_v$ can have up to $2^{\Delta}$ members. So, if graphs
with maximum degree at most $b\log n$ are considered as input to the algorithm (where $b$ is some
constant), then, for each vertex $v$ of the input graph, the size of ${\cal S}_v$ is at most
$2^{b\log n}=n^b$. So, $|{\cal S}_v|$ is polynomially bounded by the number of vertices of the
input graph $n$. Also, all other sets considered in lemma~\ref{lsetsize} are polynomially bounded
by $n$. So, for $t=3$ and for every $b>0$, the algorithm runs in polynomial time and it is efficient, when graphs $G$
with degrees less than $b\log |V(G)|$ are examined.

As mentioned in the introduction, the problem had been solved for $t=2$ on general graphs.
Now, for $t>3$, a tree in ${\cal S}_v$ may contain vertices at distance 2 from $v$. This makes the
size of ${\cal S}_v$ super-polynomial in $n$ in the worst case, when graphs with
maximum degree at most $b\log n$ are considered. Therefore, the algorithm is not efficient in
this case.

There is some possibility, though, that the $t=4$ case is similar to the $t=3$ case.
The diameter of initial partial solutions (members of ${\cal S}_v$, for each vertex $v$ of the input graph $G$; line~\ref{l1-5}
of table~\ref{ta}) is at most 2, when $t=3$, while it is at most 4, when $4\leq t\leq 5$.
One can well consider initial partial solutions of diameter at most 3, when $t=4$.
Then\footnote{A diameter at most 3 subtree of a graph $G$ consists of at most one central edge $e$ and at most
$2\Delta(G)$ edges sharing a vertex with $e$. Therefore, the number of such subtrees with a given central edge is at most
$2^{2\Delta(G)}$.}, the tree 4-spanner admissibility of graphs with degrees less than $b\log n$ (where $b$ is a
constant) may be decided efficiently too.

As mentioned in the proof of correctness (see figure~\ref{fepanalipsi}), the $t=3$ case exhibits
some structural differences as well, compared to the $t>3$ cases. These differences may justify
further investigation, in an attempt to resolve the complexity status of the tree 3-spanner
problem.

\section{Notes}
Let us hint at the diversity of the tree spanners that the algorithm can produce, with a possible application.
Assume that a tree $t$-spanner of a graph $G$ is needed\footnote{For example, a tree spanner of the
underlying graph of a communication network is needed for broadcasting but some connections (edges of the graph) are
very reliable and they must be in the spanner.} but it must contain certain edges of $G$.
Assume that these necessary edges form a tree $A$ of diameter at most $2\lfloor\frac{t}{2}\rfloor$.
Hence, there is a vertex $a$ of $G$ that is a $\lfloor\frac{t}{2}\rfloor$-center of $A$. Because of its small diameter,
$A$ is a tree $t$-spanner of $G[A]$. So, $A$ will be in ${\cal S}_a$, when algorithm {\tt Find\_Tree\_spanner}
is run on input $(G,t)$. To output only a suitable spanner, line~\ref{l1-9} of table~\ref{ta} must be changed to
\begin{equation*}
\text{{\tt{\bf If} ($V(G)=V(T^k_{v,S})$ and $S=A$) {\bf Return}($T^k_{v,S}$){\bf \}\}}}}
\end{equation*}
Then, the algorithm outputs a graph if and only if there is a tree $t$-spanner of $G$
that contains $A$.

The only reason that the algorithm is not efficient for general graphs is the huge size of sets ${\cal S}_v$ even for
a few vertices $v$ of the input graph $G$. A promising research direction is to consider a family of
input graphs for which one can prune these sets down to manageable sizes; i.e. when the algorithm
constructs set ${\cal S}_v$ for each vertex $v$ of input graph $G$ (line~\ref{l1-5} of table~\ref{ta}),
an efficient procedure may rule out many subtrees of $G$ that are not needed to build a final solution, because of
some properties of $G$.

\bibliographystyle{plain}
\bibliography{tspanners}
\end{document}